\documentclass[12pt,preprint]{article}
\usepackage{feynmf}
\usepackage{amsmath}
\usepackage{graphics}
\usepackage{epsfig}
\usepackage{psfig}
\renewcommand{\vec}[1]{{\bf #1}}
\normalsize
\newcommand{\eqb}{\begin{equation}}
\newcommand{\eqe}{\end{equation}}
\newcommand{\dmb}{\begin{displaymath}}
\newcommand{\dme}{\end{displaymath}}

\newcommand{\eab}{\begin{eqnarray}}
\newcommand{\eae}{\end{eqnarray}}

\newcommand{\be}{\begin{equation}}
\newcommand{\ee}{\end{equation}}

\begin{document}
\begin{titlepage}
\begin{flushright}
2004-10 \\
HD-THEP-10-39\\ 
\end{flushright}
\vspace{0.6cm}

\begin{center}
\Large{{\bf SU(2) Yang-Mills thermodynamics: two-loop corrections to the pressure}}

\vspace{1cm}

U. Herbst, R. Hofmann$^\dagger$, and J. Rohrer

\end{center}
\vspace{0.3cm}

\begin{center}
{\em 
Institut f\"ur Theoretische Physik\\ 
Universit\"at Heidelberg\\ 
Philosophenweg 16\\ 
69120 Heidelberg, Germany}
\end{center}
\vspace{0.3cm}
\begin{center}
{\em $\mbox{}^\dagger$Institut f\"ur Theoretische Physik\\ 
Universit\"at Frankfurt\\ 
Johann Wolfgang Goethe - Universit\"at\\ 
Robert-Mayer-Str. 10\\ 
60054 Frankfurt, Germany}
\end{center}
\vspace{0.5cm}
\newpage 

\begin{abstract}

We compute the two-loop corrections to the 
thermodynamical pressure of an SU(2) Yang-Mills theory being in 
its electric phase. Our results prove 
that the one-loop evolution of the effective gauge 
coupling constant is reliable for any 
practical purpose. We thus establish the 
validity of the picture of almost 
noninteracting thermal quasiparticles 
in the electric phase. Implications of our results 
for the explanation of the large-angle 
anomaly in the power spectrum of temperature fluctuations in the 
cosmic microwave background are discussed.

\end{abstract} 

\end{titlepage}

\section{Introduction}

In \cite{Hofmann2004} one of us has put forward an analytical and nonperturbative 
approach to the thermodynamics of SU(N) Yang-Mills theory. 
This approach self-consistently assumes the 'condensation' of (embedded) SU(2) 
trivial-holonomy calorons \cite{HarrigtonShepard1977} into a macroscopically 
stabilized adjoint Higgs field in the deconfining high-temperature phase of
the theory \footnote{This phase is referred to as electric phase in \cite{Hofmann2004}.}. 
This assumption is subject to proof which we establish 
in an analytical way in \cite{HerbstHofmann2004}. The incorporation of nontrivial-holonomy calorons 
\cite{Nahm1984,KraanVanBaalNPB1998,vanBaalKraalPLB1998,Brower1998} into the ground-state dynamics 
can be thermodynamically achieved in an exact way in terms of a 
macroscopic pure-gauge configuration. We thus describe the effects of 
dissociating nontrivial-holonomy calorons (magnetic monopoles which attract or 
repulse one another for small or large holonomy, respectively \cite{Diakonov2004}, where 
the former possibility is far more likely.) on the pressure and the energy density of the 
ground state in an average fashion, that is, thermodynamically. By a global $Z_{2,\tiny\mbox{elec}}$ 
degeneracy of the ground state and a nonvanishing expectation value of the 
Polyakov loop it can be shown analytically that the electric phase is 
deconfining. Moreover, the inrafred problem of thermal perturbation theory is resolved 
by a nontrivial ground-state structure giving rise to a mass for gauge-field 
fluctuations off the unbroken Cartan subalgebra.  

On tree-level, excitations in the electric phase are either thermal quasiparticles or 
massless 'photons'. The evolution equation for the 
effective gauge coupling $e$ in the electric phase is derived from thermodynamical 
self-consistency \cite{Gorenstein1995} which just expresses the demand 
that Legendre transformations between thermodynamical quantities, as they are derived from 
the partition function of the underlying theory, are not 
affected within the effective theory. In \cite{Hofmann2004} we have assumed a one-loop expression for
the pressure to derive the evolution $e(T)$. The purpose of this paper is to show  
that the one-loop evolution is exact for many practical purposes, that is, 
(thermal (quasi)particle) excitations in the electric phase are 
almost noninteracting throughout that phase\footnote{Some interesting 
physics does, however, take place shortly before the theory settles 
into its magnetic phase \cite{Hofmann2004}. We discuss its implications 
for the large-angle `anomaly' in the power spectrum of 
the cosmic microwave background in the last section of the present paper.}. 

The paper is organized as follows: In Sec.\,\ref{rules} we set 
up the real-time-formalism Feynman rules in unitary-Coulomb gauge 
and some notational conventions useful for organizing our calculations. 
In Sec.\,\ref{diag} we sort out the diagrams that do contribute 
to the two-loop pressure for the SU(2) case. We discuss their 
general analytical form and defere hard-core 
analytical expressions to an appendix. Kinematical constraints on 
the off-shellness of quantum fluctuations as well 
as the center-of-mass energy entering a four-gauge bosons 
vertex are being set up and discussed. In the absence of external probes to the thermalized system 
these constraints derive from the existence of a compositeness scale 
characterizing the thermodynamics of the ground state. 
In Sec.\,\ref{processing} we perform an analytical processing of the integrals 
associated with nonvanishing two-loop contributions to the pressure. 
In Sec.\,\ref{numprocessing} we discuss the problems inherent to a numerical evaluation of 
loop integrals and their solutions. For the vacuum propagation integrals are either evaluated 
in a Euclidean rotated way and a subsequent imposition of the kinematical constraints 
or by performing $\epsilon\to 0$ limits numerically in the 
Minkowskian expressions. In Sec.\,\ref{res} we present our results 
graphically. In Sec.\,\ref{SO} we discuss and summarize our work 
and point towards its possible phenomenological importance for the explanation of 
the large-angle anomaly observed in the CMB 
power spectrum \cite{WMAP2003}.

\section{Feynman rules and notational conventions\label{rules}}

\begin{figure}
\begin{center}
\includegraphics{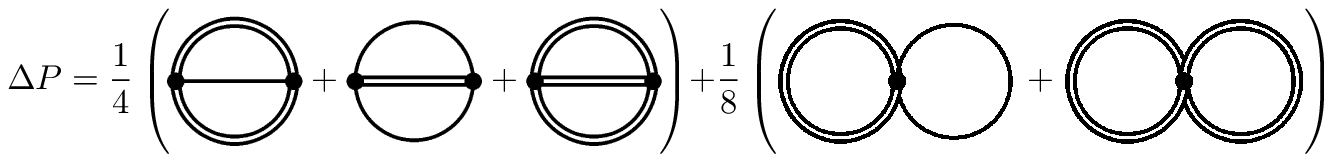}
\end{center}
\caption{Diagrams contributing to the pressure at two-loop level in a thermalized 
SU(N) Yang-Mills theory.\label{fig:contribdiagr}}      
\end{figure}

In this section we set up prerequisites for our calculations. 
The two-loop diagrams for the thermodynamical pressure split into the contributions as displayed in 
Fig.\,\ref{fig:contribdiagr}. There are local and non-local contributions. We will 
evaluate them within the real-time formalism of finite-temperature 
field theory \cite{Landsman}. For an SU(N) Yang-Mills theory 
the following rules apply:
\begin{enumerate}
\item Each diagram is divided by a factor $iV$, where $V$ denotes the number 
of vertices.
\item Local diagrams are multiplied by a factor $1/8$, 
nonlocal diagrams by $1/4$.
\end{enumerate}
The three- and four-gauge-boson vertices 
$\Gamma_{[3]abc}^{\mu\nu\rho}(p,k,q)$ and $\Gamma_{[4]abcd}^{\mu\nu\rho\sigma}$ are, respectively 
(see Fig.\ref{fig:vertices}):
\begin{figure}
\begin{center}
\includegraphics{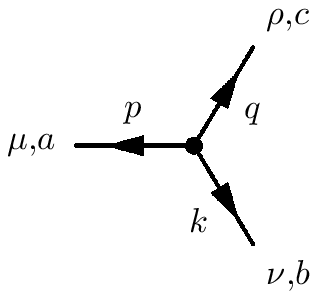}
\hspace{2.5cm}
\includegraphics{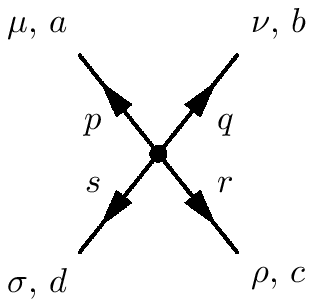}
\end{center}
\caption{The vertices $\Gamma_{[3]abc}^{\mu\nu\rho}(p,k,q)$ and $\Gamma_{[4]abcd}^{\mu\nu\rho\sigma}.$
\label{fig:vertices}}      
\end{figure}
\begin{eqnarray}
\Gamma_{[3]abc}^{\mu\nu\rho}(p,k,q)&\equiv &e\,f_{abc}[g^{\mu\nu}(p-k)^{\rho}+g^{\nu\rho}(k-q)^{\mu}
+g^{\rho\mu}(q-p)^{\nu}]\nonumber\\
\Gamma_{[4]abcd}^{\mu\nu\rho\sigma}&\equiv &-ie^2[f_{abe}f_{cde}(g^{\mu\rho}g^{\nu\sigma}-g^{\mu\sigma}g^{\nu\rho})+\nonumber\\
&&f_{ace}f_{bde}(g^{\mu\nu}g^{\rho\sigma}-g^{\mu\sigma}g^{\nu\rho})+\nonumber\\
&&f_{ade}f_{bce}(g^{\mu\nu}g^{\rho\sigma}-g^{\mu\rho}g^{\nu\sigma})]\label{eqn:vertices}
\end{eqnarray}
Since the effective theory has a stabilized\footnote{The field $\phi$ is shown 
to {\sl not} fluctuate statistically and quantum mechanically \cite{Hofmann2004}.}, 
composite, and adjoint Higgs 
field $\phi$ characterizing 
its ground state, we shall work in unitary gauge 
where $\phi$ is diagonal and the pure-gauge background is zero (see 
\cite{Hofmann2004} for a thorough discussion of the admissibility of this gauge condition). 
There is a residual gauge freedom for 
the unbroken abelian subgroup\footnote{This assumes maximal breaking by $\phi$.} U(1)$^{\tiny\mbox{N-1}}$. 
A physical gauge choice is Coulomb gauge. In unitary-Coulomb gauge 
each of the propagators for Tree-Level-Heavy/Massless (TLH/TLM) 
modes split into a vacuum and a thermal part as follows \cite{Hofmann2004,Landsman}:
\begin{eqnarray}
\label{eqn:propagators}
D_{\mu\nu,ab}^{TLH}(p)&=&-\delta_{ab}\tilde{D}_{\mu\nu}(p)
\left[\frac{i}{p^2-m^2}+2\pi\delta(p^2-m^2)n_B(|p_0/T|)\right]\\
\tilde{D}_{\mu\nu}(p)&=&\left(g_{\mu\nu}-\frac{p_{\mu}p_{\nu}}{m^2}\right)\nonumber\\
D_{\mu\nu,ab}^{TLM}(p)&=&-\delta_{ab}\bar{D}_{\mu\nu}(p)
\left[\frac{i}{p^2}+2\pi\delta(p^2)n_B(|p_0/T|)\right]\nonumber\\
\bar{D}_{\mu\nu}(p)&=&\left\{
\begin{array}[2]{cc}0&\mbox{if}\ \ \mu=0\ \ \mbox{or}\ \ \nu=0\\
\left(\delta_{\mu\nu}-\frac{p_\mu p_\nu}{\vec{p}^2}\right)&\mbox{else}\nonumber 
\end{array}\right.\,. 
\end{eqnarray}
In Eq.\,(\ref{eqn:propagators}) $n_B(x)=1/(e^{x}-1)$ denotes the Bose-Einstein 
distribution function, and $T$ is the temperature. We have neglected
the propagation of the $A_0$ field in the TLM propagator since we expect
that this field is strongly screened - for $T>2\,T_c$, 
where the TLH mass is sizably smaller than $T$, the Debye mass is $\sim eT$ 
with $e\sim 5.1$, for $T\sim T_c$ there is an exponential 
suppression of this screening. 

With these rules at hand the 
two-loop correction to the pressure is given as
\begin{eqnarray}
\Delta P=\frac{1}{8}\Delta P_{\tiny\mbox{local}}+\frac{1}{4}\Delta P_{\tiny\mbox{nonlocal}}\label{eqn:DP2total}\ 
\end{eqnarray}
where the local contributions can be written as
\begin{eqnarray}
\Delta P_{\tiny\mbox{local}}=\frac{1}{i}\int\frac{d^4p\,d^4k}{(2\pi)^8}\,\Gamma_{[4]abcd}^{\mu\nu\rho\sigma}D_{\mu\nu,ab}(p)
D_{\rho\sigma,cd}(k)\,. \label{eqn:DP2local}
\end{eqnarray}
For nonlocal diagrams we have 
\begin{eqnarray}
\Delta P_{\tiny\mbox{nonlocal}}&=&\frac{1}{2i}\int\frac{d^4p\,d^4k}{(2\pi)^8}\,\Gamma_{[3]abc}^{\lambda\mu\nu}(p,k,-p-k)
\Gamma_{[3]rst}^{\rho\sigma\tau}(-p,-k,p+k)\times\nonumber\\
&&\quad\quad D_{\lambda\rho,ar}(p)D_{\mu\sigma,bs}(k)D_{\nu\tau,ct}(-p-k)\,.\label{eqn:DP2nonlocal}
\end{eqnarray}
In Eqs.(\ref{eqn:DP2local}) and (\ref{eqn:DP2nonlocal}) 
$D_{\mu\nu,ab}$ stands for both TLH- and TLM-propagators, and one 
has to sum over all combinations allowed by the vertices $\Gamma_{[3]abc}^{\lambda\mu\nu}$ and 
$\Gamma_{[4]abcd}^{\mu\nu\rho\sigma}$.

Let us now introduce a useful convention: Due to the split of propagators 
into vacuum and thermal contributions 
in Eq.(\ref{eqn:propagators}) combinations of thermal and vacuum
contributions of TLH and TLM propagators arise in Eqs.(\ref{eqn:DP2local}) and (\ref{eqn:DP2nonlocal}). 
We will consider these contributions separately and denote them by 
\begin{eqnarray}
\Delta P^{XYZ/XY}_{\alpha_X\beta_Y\gamma_Z/\alpha_X\beta_Y}
\end{eqnarray}
where capital roman letters take the values $H$ or $M$, 
indicating the propagator type (TLH/TLM), and the associated small greek 
letters take the values $v$ (vacuum) or $t$ (thermal).

\section{Contributing diagrams for SU(2)\label{diag}}

In what follows we only investigate the case SU(2). It is clear that not 
all combinations of TLH- and TLM-propagators may contribute. 
This is due to the structure constants entering the vertices. For SU(2) they are 
$f_{abc}=\epsilon_{abc}$. As a consequence, the thirteen (naively) 
nonvanishing diagrams are 
\begin{eqnarray}
\Delta P^{HH}&=&\Delta P^{HH}_{vv}+\Delta P^{HH}_{vt}+\Delta P^{HH}_{tt}\nonumber\\
\Delta P^{HM}&=&\Delta P^{HM}_{vv}+\Delta P^{HM}_{vt}+\Delta P^{HM}_{tv}+\Delta P^{HM}_{tt}\nonumber\\
\Delta P_{HHM}&=&\Delta P^{HHM}_{vvv}+\Delta P^{HHM}_{vvt}+\Delta P^{HHM}_{tvv}+\nonumber\\
&&\Delta P^{HHM}_{ttv}+\Delta P^{HHM}_{ttt}+\Delta P^{HHM}_{vtt}\,.\label{eqn:contribdiagrams}
\end{eqnarray}
The number of allowed diagrams reduces further 
if one considers the strong coupling limit for the effective gauge coupling $e$ 
($e>0.5$). This can be seen by virtue of 
the following compositeness constraint \cite{Hofmann2004}:
\begin{eqnarray}
|p^2-m^2|\leq |\phi|^2 & \mbox{or} & p_E^2+m^2 \leq |\phi|^2\label{eqn:offshellcutoff}
\end{eqnarray}
where the index $E$ stands for the 
Euclidean rotated momentum. Eq.\,(\ref{eqn:offshellcutoff}) expresses the fact 
that the ground-state physics is characterized by a scale 
set by $|\phi|$ which determines the maximal hardness for the 
off-shellness of gauge-boson fluctuations the ground state 
can possibly generate. A Gaussian smearing of this constraint for Euclidean 
momenta introduces a ridiculously small effect since the 
variance for this distribution is, again, given by $|\phi|^2$. Notice 
that only in unitary-Coulomb gauge, that is, the only physical gauge, it makes sense to 
impose the constraint (\ref{eqn:offshellcutoff}).

By the adjoint 
Higgs mechanism the (degenerate) mass of the two TLH modes is given as \cite{Hofmann2004}
\eqb
\label{massTLH}
m=2e|\phi|
\eqe
where
\eqb
\label{modphi} 
|\phi|^2=\frac{\Lambda^3}{2\pi T}\,.
\eqe
In Eq.\,(\ref{modphi}) $\Lambda$ denotes the 
Yang-Mills scale. For later use we introduce a 
dimensionless temperature $\lambda$ as  
\eqb
\label{dimlessT} 
\lambda=\frac{2\pi T}{\Lambda}\,.
\eqe
From the one-loop evolution of the effective gauge coupling it follows that $e$ runs 
into a logarithmic pole
\eqb
\label{logpole}
e(\lambda)\sim -\log(\lambda-\lambda_c)
\eqe
at $\lambda_c=11.65$ \cite{Hofmann2004}. This is the point where the theory undergoes a 
2$^{\tiny\mbox{nd}}$ order phase transition by the condensation of magnetic monopoles, 
and thus $\lambda_c$ corresponds to the lowest attainable temperature in the electric phase.   

We can scale out $|\phi|$ in Eq.\,(\ref{eqn:offshellcutoff}). 
Then the Euclidean constraint becomes
\eqb
\label{dimlessconstraint}
\sqrt{w^2+(2e)^2}\leq 1\,
\eqe
where $w^2\equiv p_E^2/|\phi|^2$. Since $w^2$ is always positive we conclude that 
only for $e\leq 0.5$ we do get a contribution from TLH vacuum fluctations in loop 
integrals. The plateau-value\footnote{This plateau indicates the conservation of isolated magnetic 
charge for monopoles contributing to the ground-state thermodynamics. It is an attractor 
of the (downward) evolution signalling the UV-IR decoupling property 
that follows from the renormalizability of the underlying theory.} 
for $e$ is, however, $e\sim 5.1$ as a result of the one-loop evolution \cite{Hofmann2004}. TLM vacuum 
modes do contribute, however, and we are left with the computation of 
$\Delta P^{HH}_{tt}$, $\Delta P^{HM}_{tt}$, $\Delta P^{HM}_{tv}$ 
and $\Delta P^{HHM}_{ttv}$ ($\Delta P^{HHM}_{ttt}$ vanishes by momentum conservation).

There is one more kinematical constraint: 
For a thermalized system with no external probes applied to it, 
the center-of-mass energy flowing into a four-vertex 
must not be greater than the compositeness scale $|\phi|$ 
of the effective theory. That is, the hot-spot generated within the 
vertex must not destroy the ground state 
of the system locally since the modes entering the vertex were generated 
by the very same ground state elsewhere. This is expressed as
\begin{eqnarray}
|\phi|^2\geq |(p+k)^2|\label{eqn:integrationrestriction}
\end{eqnarray}
where $p$ and $k$ are the momenta of the modes entering the vertex. As we shall see, 
Eq.\,(\ref{eqn:integrationrestriction}) leads to a strong restriction in the loop integration. 

To perform the contractions in Eqs.\,(\ref{eqn:DP2local}) and (\ref{eqn:DP2nonlocal}) 
it is useful to exploit the transversality of the tensorial part $\bar{D}_{\mu\nu}(q)$ of the
TLM propagator from the start. The following four relations hold:
\begin{eqnarray}
\bar{D}_{\mu\nu}(q)q^{\mu}&=&0\nonumber\\
\bar{D}_{\mu\nu}(q)g^{\mu\nu}&=&-2\nonumber\\
\bar{D}_{\mu\nu}(q)p^{\mu}p^{\nu}&=&|\vec{p}|^2-\frac{(\vec{q}\vec{p})^2}{|\vec{q}|^2}\nonumber\\
\bar{D}_{\mu\nu}(q)p^{\mu}k^{\nu}&=&
\vec{pk}-\frac{(\vec{k}\vec{q})(\vec{p}\vec{q})}{|\vec{q}|^2}\,.\label{eqn:TLMfeatures}
\end{eqnarray}
The results for all relevant contractions are derived in the Appendix. 

\section{Calculation of the integrals\label{processing}}

With the contractions of tensor structures at hand, we are now in a position to calculate all 
two-loop corrections. For $\Delta P^{HH}_{tt}$ this is done in detail, 
for the other contributions we resort to a more 
compact presentation. We have
\begin{eqnarray}
\label{HHtt richtig}
\Delta P^{HH}_{tt}&=&\frac{1}{i}\int\frac{d^4p\,d^4k}{(2\pi)^8}\,
\Gamma_{[4]aacc}^{\mu\nu\rho\sigma}\tilde{D}_{\mu\nu}(p)\tilde{D}_{\rho\sigma}(k)\times\nonumber\\
&&(2\pi)\delta(p^2-m^2)\,n_B(|p_0/T|)(2\pi)\delta(k^2-m^2)\,n_B(|k_0/T|)\nonumber\\
&=&-2e^2\int\frac{d^4p\,d^4k}{(2\pi)^6}\left(24-6\frac{p^2}{m^2}-6\frac{k^2}{m^2}+2\frac{p^2k^2}{m^4}
-2\frac{(pk)^2}{m^4}\right)\times\nonumber\\
&&\delta(p^2-m^2)n_B(|p_0/T|)\,\delta(k^2-m^2)n_B(|k_0/T|)\,.
\end{eqnarray}
In Eq.\,(\ref{HHtt richtig}) both color indices $a,c$ are summed over $a,c=1,2$. 
The product of $\delta$-functions can be rewritten as
\begin{eqnarray*}
\delta(p^2-m^2)\delta(k^2-m^2)&=&\frac{1}{4\sqrt{\vec{p}^2+m^2}\sqrt{\vec{k}^2+m^2}}\times\\ 
&&\big[\delta(p_0-\sqrt{\vec{p}^2+m^2})\delta(k_0-\sqrt{\vec{k}^2+m^2})+\\
&&\delta(p_0-\sqrt{\vec{p}^2+m^2})\delta(k_0+\sqrt{\vec{k}^2+m^2})+\\
&&\delta(p_0+\sqrt{\vec{p}^2+m^2})\delta(k_0-\sqrt{\vec{k}^2+m^2})+\\
&&\delta(p_0+\sqrt{\vec{p}^2+m^2})\delta(k_0+\sqrt{\vec{k}^2+m^2})\big]\,.
\end{eqnarray*}
The contraction $\Gamma_{[4]aacc}^{\mu\nu\rho\sigma}\tilde{D}_{\mu\nu}(p)\tilde{D}_{\rho\sigma}(k)$
contains only even products of $k$ and $p$ (this is also true for the other contractions), 
like $p^2$, $k^2$ or $pk$. Thus, performing the zero-component 
integration over either
\begin{eqnarray*}
&\delta(p_0-\sqrt{\vec{p}^2+m^2})\,\delta(k_0+\sqrt{\vec{k}^2+m^2})&\\
&\mbox{or}&\\
&\delta(p_0+\sqrt{\vec{p}^2+m^2})\,\delta(k_0-\sqrt{\vec{k}^2+m^2})\,,&
\end{eqnarray*}
(signs in the argument of $\delta$-functions opposite, crossterms) leads to the same result. 
This is also true for the two uncrossed products of $\delta$-functions with equal signs. 
After the integration is performed we may therefore set
\begin{eqnarray}
p^2&\rightarrow& m^2\nonumber\\
k^2&\rightarrow& m^2\nonumber\\
(pk)&\rightarrow& \pm\sqrt{\vec{p}^2+m^2}\sqrt{\vec{k}^2+m^2}-\vec{p}\vec{k}\nonumber\\
(pk)^2&\rightarrow& \vec{p}^2\vec{k}^2+(\vec{p}^2+\vec{k}^2)m^2+
m^4\mp 2\vec{p}\vec{k}\sqrt{\vec{p}^2+m^2}\sqrt{\vec{k}^2+m^2}
+(\vec{p}\vec{k})^2\,.\nonumber\\ \label{eqn:0integration}
\end{eqnarray}
The upper case is obtained when the signs are equal, the lower case when they are opposite.

Examining the integration constraint in Eq.\,(\ref{eqn:integrationrestriction}) 
after the zero-component integration over the products of $\delta$-functions 
is performed shows that only the combinations with 
opposite signs must be evaluated: 
\begin{eqnarray}
|\phi|^2&\geq&|(p+k)^2|=|p_{0}^2-\vec{p}^2+k_{0}^2-\vec{k}^2+2p_0k_0-2\vec{p}\vec{k}|\nonumber\\
&\rightarrow& |\phi|^2\geq |2m^2\pm 2\sqrt{\vec{p}^2+m^2}\sqrt{\vec{k}^2+m^2}-2\vec{p}\vec{k}|\nonumber\\
&\rightarrow& 1\geq 2|(2e)^2\pm \sqrt{x^2+(2e)^2}
\sqrt{y^2+(2e)^2}-xy\cos{\theta}|
\label{eqn:HHttrestriction}
\end{eqnarray}
where we have introduced (also for later use) dimensionless variables
\begin{eqnarray}
x=|\vec{p}|/|\phi|\,,&&y=|\vec{k}|/|\phi|\,,\nonumber\\
z=\cos{\theta}\,,&&\lambda^{-3/2}=\frac{|\phi|}{2\pi T}\label{eqn:scaling}\,.
\end{eqnarray}
In Eq.\,(\ref{eqn:HHttrestriction}) $\theta$ denotes the angle between ${\bf p}$ and ${\bf k}$. 
We observe that for the "$+$" case the difference 
between the second and third term is always positive. And, 
because of the first term, the whole expression is greater than unity in the 
strong coupling limit. Thus only the "$-$" case needs to be considered. 
This is also true for $\Delta P^{HM}_{tt}$ and $\Delta P^{HHM}_{ttv}$ 
though the analytical expressions may look different.\\ 

Applying the knowledge gathered in the Appendix, $\Delta P^{HH}_{tt}$ can be reduced to
\begin{eqnarray*} 
\Delta P^{HH}_{tt}&=&-2e^2\int\frac{d^4p\,d^4k}{(2\pi)^6}\left(24-6\frac{p^2}{m^2}-6\frac{k^2}{m^2}+2\frac{p^2k^2}{m^4}-2\frac{(pk)^2}{m^4}\right)\times\\
&&\delta(p_0-\sqrt{\vec{p}^2+m^2})
\delta(k_0+\sqrt{\vec{k}^2+m^2})\frac{n_B(|p_0/T|)n_B(|k_0/T|)}
{2\sqrt{\vec{p}^2+m^2}\sqrt{\vec{k}^2+m^2}}\,.
\end{eqnarray*}
Integrating over the zero components by using Eq.\,(\ref{eqn:0integration}), we arrive at 
\begin{eqnarray*}
\Delta P^{HH}_{tt}&=&-2e^2\int\frac{d^3\vec{p}\,d^3\vec{k}}{(2\pi)^6}\,
\frac{n_B(\sqrt{\vec{p}^2+m^2}/T)n_B(\sqrt{\vec{k}^2+m^2}/T)}{2\sqrt{\vec{p}^2+m^2}\sqrt{\vec{k}^2+m^2}}\times\\
&&\Big[12-2\frac{\vec{p}^2}{m^2}-2\frac{\vec{k}^2}{m^2}-
2\frac{\vec{p}^2\vec{k}^2}{m^4}
-2\frac{(\vec{p}\vec{k})^2}{m^4}-4\frac{\vec{p}\vec{k}}{m^4}
\sqrt{\vec{p}^2+m^2}\sqrt{\vec{k}^2+m^2}\Big]\,.
\end{eqnarray*}
After a change to polar coordinates and an evaluation of the angular integrals 
the remaining integration measure takes the form 
$2(2\pi)^2|\vec{p}|^2|\vec{k}|^2d|\vec{p}|d|\vec{k}|d\cos{\theta}$.
As a last step we re-scale variables according to Eq.\,(\ref{eqn:scaling}). 
This re-casts the kinematic constraints of Eq.\,(\ref{eqn:HHttrestriction}) into 
the following form:
\begin{eqnarray}
-1/2\leq & (2e)^2-\sqrt{x^2+(2e)^2}\sqrt{y^2+(2e)^2}-xyz & \leq +1/2\,.\label{eqn:HHrestriction}
\end{eqnarray}
Our final result for $\Delta P^{HH}_{tt}$ reads:\\ 
{\bf TLH-TLH-thermal-thermal:}
\begin{eqnarray}
\Delta P^{HH}_{tt}&=&\frac{-2e^2T^4}{\lambda^6}\int dx\, dy\, dz\;
\frac{x^2y^2}{\sqrt{x^2+(2e)^2}\sqrt{y^2+(2e)^2}}\times\nonumber\\
&&\Big[12-2\frac{x^2}{(2e)^2}-2\frac{y^2}{(2e)^2}-2\frac{x^2y^2}{(2e)^4}-2\frac{x^2y^2z^2}{(2e)^4}-
4\frac{xyz}{(2e)^4}\sqrt{x^2+(2e)^2}\sqrt{y^2+(2e)^2}\Big]\times\nonumber\\
&&n_B(2\pi\lambda^{-3/2}\sqrt{x^2+(2e)^2})
n_B(2\pi\lambda^{-3/2}\sqrt{y^2+(2e)^2})\label{eqn:P2HHtt}\,
\end{eqnarray}
where the integration is subject to the contraint in Eq.\,(\ref{eqn:HHrestriction}). 
The other two-loop corrections $\Delta P^{HM}_{tt}$, $\Delta P^{HM}_{tv}$ and $\Delta P^{HHM}_{ttv}$ 
are calculated in essentially the same way:\\ 
{\bf TLH-TLM-thermal-thermal:}\\ 
We have 
\begin{eqnarray*}
\Delta P^{HM}_{tt}&=&\frac{1}{i}\int\frac{d^4p\,d^4k}{(2\pi)^6}\,\Gamma_{[4]aa33}^{\mu\nu\rho\sigma}
\tilde{D}_{\mu\nu}(p)\bar{D}_{\rho\sigma}(k)\times\\
&&n_B(|p_0/T|)n_B(|k_0/T|)\delta(p^2-m^2)\delta(k^2)\,
\end{eqnarray*}
where the sum is over $a=1,2$. 
Consider the integration constraint Eq.\,(\ref{eqn:integrationrestriction}): 
\begin{eqnarray}
\label{cmcdl}
|\phi|^2&\geq&|(p+q)^2|=|p_0^2-\vec{p}^2+k_0^2-\vec{k}^2+2p_0k_0-2\vec{p}\vec{k}|\nonumber\\
&\rightarrow& |\phi|^2\geq|m^2\pm 2|\vec{k}|\sqrt{\vec{p}^2+m^2}-2|\vec{p}||\vec{k}|\cos{\theta}|\nonumber\\
&\rightarrow& 1\geq |(2e)^2\pm 2y\sqrt{x^2+(2e)^2}-2xy \cos{\theta}|\,.
\end{eqnarray}
Again, the "$+$" case cannot be satisfied in the strong coupling limit ($e>0.5$), 
so only the "$-$" case needs to be considered. Then $\Delta P^{HM}_{tt}$ reduces to\newpage 
\begin{eqnarray}
\Delta P^{HM}_{tt}
&=&-2e^2\int\frac{d^4p\,d^4k}{(2\pi)^6}\left(-12+4\frac{p^2}{m^2}+2\frac{\vec{p}^2\sin^2{\theta}}{m^2}\right)\times\nonumber\\
&&\frac{n_B(|p_0/T|)n_B(|k_0/T|)}{2|\vec{k}|\sqrt{\vec{p}^2+m^2}}\delta(p_0-\sqrt{\vec{p}^2+m^2})\delta(k_0+|\vec{k}|)
\nonumber\\
&=&-\frac{2e^2}{(2\pi)^4}\int d|\vec{p}|\,d|\vec{k}|\,d(\cos{\theta})\,\vec{p}^2\vec{k}^2
\left(-8+2\frac{\vec{p}^2\sin^2{\theta}}{m^2}\right)\times\nonumber\\
&&\frac{n_B(\sqrt{\vec{p}^2+m^2}/T)n_B(|\vec{k}|/T)}{|\vec{k}|\sqrt{\vec{p}^2+m^2}}\nonumber\\
&=&-\frac{2e^2T^4}{\lambda^6}\int dx\,dy\,dz\,x^2y\left(-8+2\frac{x^2(1-z^2)}{(2e)^2}\right)\times\nonumber\\
&&\frac{n_B(2\pi\lambda^{-3/2}\sqrt{x^2+(2e)^2})n_B(2\pi\lambda^{-3/2}y)}{\sqrt{x^2+(2e)^2}}
\label{eqn:P2HMtt}\,,
\end{eqnarray}
subject to the constraint Eq.\,(\ref{cmcdl}).\\ 
{\bf TLH-TLM-thermal-vacuum:} \\ 
We have 
\begin{eqnarray*}
\Delta P^{HM}_{tv}&=&\frac{1}{i}\int\frac{d^4p\,d^4k}{(2\pi)^7}\,\Gamma_{[4]aa33}^{\mu\nu\rho\sigma}
\tilde{D}_{\mu\nu}(p)\bar{D}_{\rho\sigma}(k)\, n_B(|p_0/T|)\,\frac{i}{k^2}\,\delta(p^2-m^2)\,.
\end{eqnarray*}
After the $p_0$-integration is performed the integration constraints Eqs. (\ref{eqn:offshellcutoff}) 
and (\ref{eqn:integrationrestriction}) read:
\begin{eqnarray}
\label{oHMVT}
|k^2|\leq |\phi|^2&\rightarrow&|\gamma^2-y^2|\leq 1\\ 
\label{cHMVT}
|(p+k)^2|\leq |\phi|^2&\rightarrow &|(2e)^2+\gamma^2-y^2\pm 
2\gamma\sqrt{x^2+(2e)^2}-2xy\cos{\theta}|\leq 1\,.\nonumber\\ 
\end{eqnarray}
Thus, the $k_0$- or $\gamma$-integration ($\gamma$ is the re-scaled $k_0$-component) 
cannot be performed analytically. We have
\begin{eqnarray}
\label{PHMTVana}
\Delta P^{HM}_{tv}&=&-2ie^2\int\frac{d^4p\,d^4k}{(2\pi)^7} \left(-12+4\frac{p^2}{m^2}+2\frac{\vec{p}^2\sin^2{\theta}}{m^2}\right)\times
\nonumber\\
&&\frac{n_B(\sqrt{\vec{p}^2+m^2}/T)}{\sqrt{\vec{p}^2+m^2}}\frac{1}{k^2}\delta(p_0-\sqrt{\vec{p}^2+m^2})\nonumber\\
&=&-4ie^2\int\frac{d|\vec{p}|\,dk_0\,d|\vec{k}|\,d(\cos{\theta})}{(2\pi)^5\sqrt{\vec{p}^2+m^2}}\,
\vec{p}^2\vec{k}^2\left(-8+2\frac{\vec{p}^2\sin^2{\theta}}{m^2}\right)\times
\nonumber\\
&& n_B(\sqrt{\vec{p}^2+m^2}/T)\frac{1}{k_0^2-\vec{k}^2}\nonumber\\
&=&\frac{-4ie^2T^4}{(2\pi)^5\lambda^6}\int\frac{dx\,dy\,d\gamma\,dz}{\sqrt{x^2+(2e)^2}}\,x^2y^2
\left(-8+2\frac{x^2(1-z^2)}{(2e)^2}\right)\times
\nonumber\\
&& n_B(2\pi\lambda^{-3/2}\sqrt{x^2+(2e)^2})\frac{1}{\gamma^2-y^2}\,.
\end{eqnarray}
This is, however, not easy to evaluate numerically. To show the smallness of $\Delta P^{HM}_{tv}$ 
we resort to estimating an upper bound on the modulus of the integral in Eq.\,(\ref{PHMTVana}). This 
is done by neglecting the center-of-mass energy constraint Eq.\,(\ref{cHMVT}) 
completely (but taking into account the constraint Eq.\,(\ref{oHMVT})) and by integrating over the modulus of the integrand: 
\begin{eqnarray}
\left|\Delta P^{HM}_{tv}\right|&\leq&-2ie^2\int\frac{d^3p\,d^4k}{(2\pi)^7}\,\frac{1}{k^2}\Big|
\left(-8+2\frac{\vec{p}^2\sin^2{\theta}}{m^2}\right)\times
\frac{n_B(\sqrt{\vec{p}^2+m^2}/T)}{\sqrt{\vec{p}^2+m^2}}\Big|\nonumber\\
&=& e^2\int\frac{d|\vec{p}|\,d(\cos{\theta})\,dk}{(2\pi)^4}\,k\,\Big|\vec{p}^2
\left(-8+2\frac{\vec{p}^2\sin^2{\theta}}{m^2}\right)\times
\frac{n_B(|\sqrt{\vec{p}^2+m^2}/T|)}{\sqrt{\vec{p}^2+m^2}}\Big|\nonumber\\
&=&\frac{e^2T^4}{2\lambda^6}\int dx\,dz\,\Big|\left(-8+2\frac{x^2(1-z^2)}{(2e)^2}\right)\times
\frac{x^2}{\sqrt{x^2+(2e)^2}}\times\nonumber\\
&&\;n_B(2\pi\lambda^{-3/2}\sqrt{x^2+(2e)^2})\Big|\label{eqn:P2HMtv}\,.
\end{eqnarray}
In the second line of Eq.\,(\ref{eqn:P2HMtv}) $k$ has the meaning of $k\equiv\sqrt{k_E^2}$.
\newpage   
{\bf TLH-TLH-TLM-thermal-thermal-vacuum:}\\ 
Here we have
\begin{eqnarray*}
\Delta P^{HHM}_{ttv}&=&-\frac{1}{2i}\int\frac{d^4p\,d^4k\,d^4q}{(2\pi)^6}\,\Gamma_{[3]ab3}^{\lambda\mu\nu}(p,k,q)
\Gamma_{[3]ab3}^{\rho\sigma\tau}(-p,-k,-q)\times\\
&&\tilde{D}_{\lambda\rho}(p)\tilde{D}_{\mu\sigma}(k)\bar{D}_{\nu\tau}(q)
\, n_B(|p_0/T|)n_B(|k_0/T|)\times\\
&&\frac{i}{(k+p)^2}\delta(p^2-m^2)\delta(k^2-m^2)\delta(q+p+k)\,
\end{eqnarray*}
where the sum is over $a,b=1,2$. 
Due to momentum conservation both kinematic constraints, Eqs.\,(\ref{eqn:offshellcutoff}) and 
(\ref{eqn:integrationrestriction}), are equivalent:
\begin{eqnarray*}
|q|^2&=&|(p+k)^2|=|p^2+k^2+2pk|\leq|\phi|^2\\
&\rightarrow& |2(2e)^2\pm 2\sqrt{x^2+(2e)^2}\sqrt{y^2+(2e)^2}-2xy\cos{\theta}|\leq 1\,.
\end{eqnarray*}
This is the same as for $\Delta P^{HH}_{tt}$, so only the "$-$" case needs to be considered:
\begin{eqnarray}
\Delta P^{HHM}_{ttv}&=&-\frac{2e^2}{2i}\int\frac{d^4p\,d^4k}{(2\pi)^6}
	\Big[10p^2+10k^2+16pk-2\frac{p^4}{m^2}-2\frac{k^4}{m^2}-\nonumber\\
	&&8\frac{p^2(pk)}{m^2}-8\frac{k^2(pk)}{m^2}-16\frac{(pk)^2}{m^2}
	-\frac{\vec{p}^2\vec{k}^2\sin^2{\theta}}{(\vec{p}+\vec{k})^2}\times\nonumber\\ 
	&&\Big(10-3\frac{p^2}{m^2}-3\frac{k^2}{m^2}-8\frac{pk}{m^2}+\frac{p^4}{m^4}+\frac{k^4}{m^4}+\nonumber\\
	&&4\frac{p^2(pk)}{m^4}+4\frac{k^2(pk)}{m^4}+4\frac{(pk)^2}{m^4}+2\frac{p^2k^2}{m^4}\Big)\Big]\times\nonumber\\
	&&\frac{n_B(\sqrt{\vec{p}^2+m^2}/T)n_B(\sqrt{\vec{k}^2+m^2}/T)}
	{2\sqrt{\vec{p}^2+m^2}\sqrt{\vec{k}^2+m^2}}
	\frac{i}{(k+p)^2}\times\nonumber\\
	&&\delta(p_0-\sqrt{\vec{p}^2+m^2})\delta(k_0+\sqrt{\vec{k}^2+m^2})\,.
\end{eqnarray}
Using Eq.\,(\ref{eqn:0integration}), 
the part $\propto \frac{\vec{p}^2\vec{k}^2\sin^2{\theta}}{(\vec{p}+\vec{k})^2}$ in the square
brackets after $p_0$ and $k_0$ integration reads 
\begin{eqnarray*}
12+4\frac{\vec{p}^2}{m^2}+4\frac{\vec{k}^2}{m^2}+4\frac{\vec{p}^2\vec{k}^2(1-z^2)}{m^4}+
8\frac{|\vec{p}||\vec{k}|z}{m^4}\sqrt{\vec{p}^2+m^2}\sqrt{\vec{k}^2+m^2}
\end{eqnarray*}
where polar coordinates have already been introduced.\\
The remaining part is
\begin{eqnarray*}
-16\left[\vec{p}^2+\vec{k}^2+\frac{\vec{p}^2\vec{k}^2(1-z^2)}{m^2}+2\frac{|\vec{p}||\vec{k}|z}{m^2}
\sqrt{\vec{p}^2+m^2}\sqrt{\vec{k}^2+m^2}\right]\,.
\end{eqnarray*}
The propagator $1/(p+k)^2$ becomes after re-scaling
\begin{eqnarray*}
\frac{1}{(p+k)^2}\rightarrow \frac{1}{2(2e)^2-2\sqrt{x^2+(2e)^2}\sqrt{y^2+(2e)^2}-2xyz}\,.
\end{eqnarray*}
Thus, we have
\begin{eqnarray}
\Delta P^{HHM}_{ttv}&=&\frac{e^2T^4}{2\lambda^6}\int dx\,dy\,dz\,\frac{x^2y^2}{\sqrt{x^2+(2e)^2}\sqrt{y^2+(2e)^2}}\times\nonumber\\
&&\frac{n_B(2\pi\lambda^{-3/2}\sqrt{x^2+(2e)^2})\,n_B(2\pi\lambda^{-3/2}\sqrt{y^2+(2e)^2})}{(2e)^2-\sqrt{x^2+(2e)^2}
\sqrt{y^2+(2e)^2}-xyz}\nonumber\\
&&\Big\{16\Big[x^2+y^2+2\frac{xyz}{(2e)^2}\sqrt{x^2+(2e)^2}\sqrt{y^2+(2e)^2}+\nonumber\\
&&\frac{x^2y^2(1+z^2)}{(2e)^2}\Big]+\frac{x^2y^2(1-z^2)}{x^2+y^2+2xyz}
\Big[12+4\frac{x^2}{(2e)^2}+4\frac{y^2}{(2e)^2}+\nonumber\\
&&+4\frac{x^2y^2(1+z^2)}{(2e)^4}+8\frac{xyz}{(2e)^4}\sqrt{x^2+(2e)^2}\sqrt{y^2+(2e)^2}\Big]\Big\}\,.\label{eqn:P2HHMttv}
\end{eqnarray}

\section{Numerical integration\label{numprocessing}}
The objective of this section is to numerically evaluate the expressions 
(\ref{eqn:P2HHtt}), (\ref{eqn:P2HMtt}),(\ref{eqn:P2HMtv}), and (\ref{eqn:P2HHMttv}). 

Two observations should already be pointed out here:\\ 
(1) As it will turn out, ignoring the kinematical constraint 
Eq.\,(\ref{cHMVT}) in the 
expression for $\Delta P^{HM}_{tv}$ gives an 
upper bound which is much smaller in modulus than the by-far dominating 
contribution subject to these constraints for $\lambda$ not too far above $\lambda_c$. 
While for the former the 
exact implementation of the constraints 
is virtually impossible it is difficult but doable for the others.\\ 
(2) The nonlocal correction has a singular integrand due to the 
TLM propagator being massless.\\ 
Both problems are resolved in the following two sections. 

\subsection{Constraints on integrations}
\begin{figure}
\begin{center}
\leavevmode
\leavevmode
\vspace{5.3cm}
\includegraphics{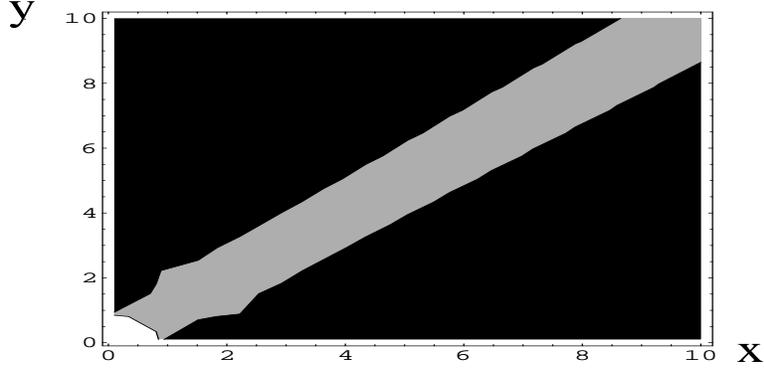}
\end{center}
\caption{Constraints on integration for $\Delta P^{HH}_{tt}$: The lower bound in $z$ 
is always $-1$ while the upper bound is not. White area: restriction is always satisfied, 
$z_{max}=1$, grey area: restriction can be satisfied, 
$z_{max}$ is given by $z_+$ as in  Eq.(\ref{eqn:contplotrestriction}), 
black area: restriction can never be satisfied.
\label{fig:contourplotHHtt}}
\end{figure}
A straight-forward implementation of the kinematical constraints is to 
multiply the integrands with appropriate $\Theta$-functions. 
This, however, cannot straight-forwardly be fed into a Mathematica program. 
Here, we demonstrate how the problem is tackled
for $\Delta P^{HH}_{tt}$.\\ 
Eq.\,(\ref{eqn:HHrestriction}) can be rewritten as
\begin{eqnarray}
z&=&\cos{\theta}\leq \frac{1+2(2e)^2-2\sqrt{x^2+(2e)^2}\sqrt{y^2+(2e)^2}}{2xy}\nonumber\\
&&\mbox{and}\nonumber\\
z&=&\cos{\theta}\geq \frac{-1+2(2e)^2-2\sqrt{x^2+(2e)^2}\sqrt{y^2+(2e)^2}}{2xy}
\end{eqnarray}
where $-1\le z\le 1$. For the lower and upper integration limit we therefore get
\begin{eqnarray}
\label{zminmax}
&&\mbox{lower limit: }z_{min}=\min[1,\max\left[-1,z_-(x,y,e)\right]]\nonumber\\
&&\mbox{upper limit: }z_{max}=\max[-1,\min\left[1,z_+(x,y,e)\right]]
\end{eqnarray}
with the definitions
\begin{eqnarray}
z_-(x,y,e)&\equiv&\frac{-1+2(2e)^2-2\sqrt{x^2+(2e)^2}\sqrt{y^2+(2e)^2}}{2xy}\nonumber\\
z_+(x,y,e)&\equiv&\frac{1+2(2e)^2-2\sqrt{x^2+(2e)^2}\sqrt{y^2+(2e)^2}}{2xy}\,.\label{eqn:contplotrestriction}
\end{eqnarray}
Notice that $z_{min}(x,y,5.1)$ always equals $-1$. A contour plot for $z_+(x,y,5.1)$ is 
displayed in Fig.\,\ref{fig:contourplotHHtt}. This plot shows that the constraint hardly ever 
is satisfied. We observe that $z_+(x,y,5.1)$ is smaller than $-1$ in the black 
area, greater than $+1$ in the white and inbetween these boundaries in the grey area. 
The integration is restricted to a small band around $x=y$ only. Parameterizing this area leads to 
an upper and lower limit for the integration range in $y=y(x)$ (depending on $x$). 
Looking at Fig.\,\ref{fig:contourplotHHtt}, 
one also sees that $x$ runs from zero to infinity. For $\Delta P^{HHM}_{ttv}$ 
the constraints are the same. 
For $\Delta P^{HM}_{tt}$  we have to re-adjust our definitions of the integration limits. 
The upper and lower limits of integration in $z$ formally are defined 
as in Eq.\,(\ref{zminmax}) with the difference that $z_{\pm}$ now are given as
\begin{eqnarray}
z_-(x,y,e)&\equiv&\frac{-1+2(2e)^2-2y\sqrt{y^2+(2e)^2}}{2xy}\nonumber\\
z_+(x,y,e)&\equiv&\frac{1+2(2e)^2-2y\sqrt{y^2+(2e)^2}}{2xy}\,.\label{alter}
\end{eqnarray}
 
\subsection{Singular integrand in the nonlocal diagram}

For $\Delta P^{HHM}_{ttv}$ an additional problem arises. Consider the integrand:
\begin{eqnarray}
&&\frac{x^2y^2}{\sqrt{x^2+(2e)^2}\sqrt{y^2+(2e)^2}}
\frac{n_B(2\pi\lambda^{-3/2}\sqrt{x^2+(2e)^2})n_B(2\pi\lambda^{-3/2}\sqrt{x^2+(2e)^2})}{(2e)^2-\sqrt{x^2+(2e)^2}\sqrt{y^2+(2e)^2}-xyz}\nonumber\\
&&\Big\{16\Big[x^2+y^2+\frac{x^2y^2(1+z^2)}{(2e)^2}+2\frac{xyz}{(2e)^2}\sqrt{x^2+(2e)^2}\sqrt{y^2+(2e)^2}\Big]\nonumber\\
&&+\frac{x^2y^2(1-z^2)}{x^2+y^2+2xyz}
\Big[12+4\frac{x^2}{(2e)^2}+4\frac{y^2}{(2e)^2}+4\frac{x^2y^2(1-z^2)}{(2e)^4}\nonumber\\
&&+8\frac{xyz}{(2e)^4}\sqrt{x^2+(2e)^2}\sqrt{y^2+(2e)^2}\Big]\Big\}\,.
\end{eqnarray}
The first part in curly brackets has no singularity and can be integrated numerically 
without additional thinking. The part $\propto \frac{x^2y^2(1-z^2)}{x^2+y^2+2xyz}$ can not 
be integrated numerically as it stands since 
it diverges at $x=y$ and $z=-1$. Complex analysis, that is, the residue theorem, can not be applied to this problem 
because we can not close the line integral at infinity due to the integration constraint. 
We therefore add $i\epsilon\,\ (\epsilon>0)$ to the inverse TLM propagator. 
One needs to prescribe a 
small value for $\epsilon$ and check the numerical convergence of the integral 
in the limit $\epsilon\rightarrow 0$. The results for $\lambda=70,200$ are shown in 
Table\,\ref{tab:epsilon}: The real part stabilizes while the imaginary part 
converges to zero. In our computations a value $\epsilon=10^{-7}$ is 
reasonable in view of available numerical precision. 
\begin{table}
\begin{center}
\begin{tabular}{|l||r|r|r|r|}
\hline
$\epsilon$&$10^{-6}$&$10^{-7}$&$10^{-8}$&$10^{-9}$\\ \hline
$T^{-4}\times$ Re $\Delta P^{HHM}_{ttv}(70)$($\times-10^{-2}$)&$2.1028$&$2.1051$&$2.1058$&$2.1060$ \\ \hline
$T^{-4}\times$ Im $\Delta P^{HHM}_{ttv}(70)$&$-3\times 10^{-5}$&$-1\times 10^{-5}$&$-3\times 10^{-6}$&$-3\times 10^{-6}$  \\ \hline
$T^{-4}\times$ Re $\Delta P^{HHM}_{ttv}(200)$($\times -10^{-3}$)&$9.8744$&$9.8850$&$9.8882$&$9.8892$ \\ \hline
$T^{-4}\times$ Im $\Delta P^{HHM}_{ttv}(200)$&$-1\times  10^{-5}$&$-5\times 10^{-6}$&$-1\times 10^{-6}$&$-4\times 10^{-7}$\\ \hline  
\end{tabular}
\label{tab:epsilon}
\caption{Numerical evaluation of the $\epsilon$-dependent 
part of $\Delta P^{HHM}_{ttv}$ (adding a term $i\epsilon$ ($\epsilon>0$) to the 
inverse TLM propagator in Minkowskian signature). Obviously, 
the real part of the integral is not sensitive to the value of 
$\epsilon$ while the imaginary part tends to zero for $\epsilon\rightarrow 0$. }
\end{center}
\end{table}

\section{Results\label{res}}

\begin{figure}
\begin{center}
\leavevmode
\leavevmode
\vspace{4.3cm}
\includegraphics{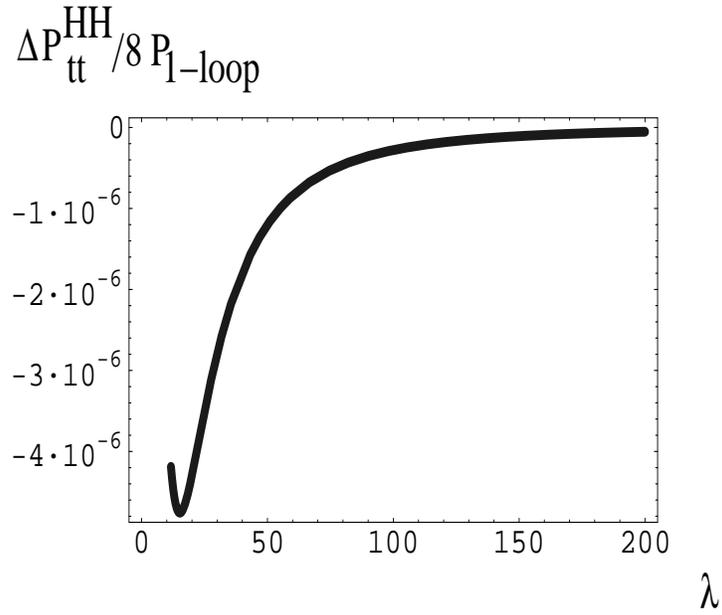}
\end{center}
\caption{Ratio of $\frac{1}{8}\Delta P^{HH}_{tt}$ and $P_{\tiny\mbox{1-loop}}$ as a function of $11.65\le\lambda\le 200$.\label{Fig-4.ps}}
\end{figure}

\begin{figure}
\begin{center}
\leavevmode
\leavevmode
\vspace{4.3cm}
\includegraphics{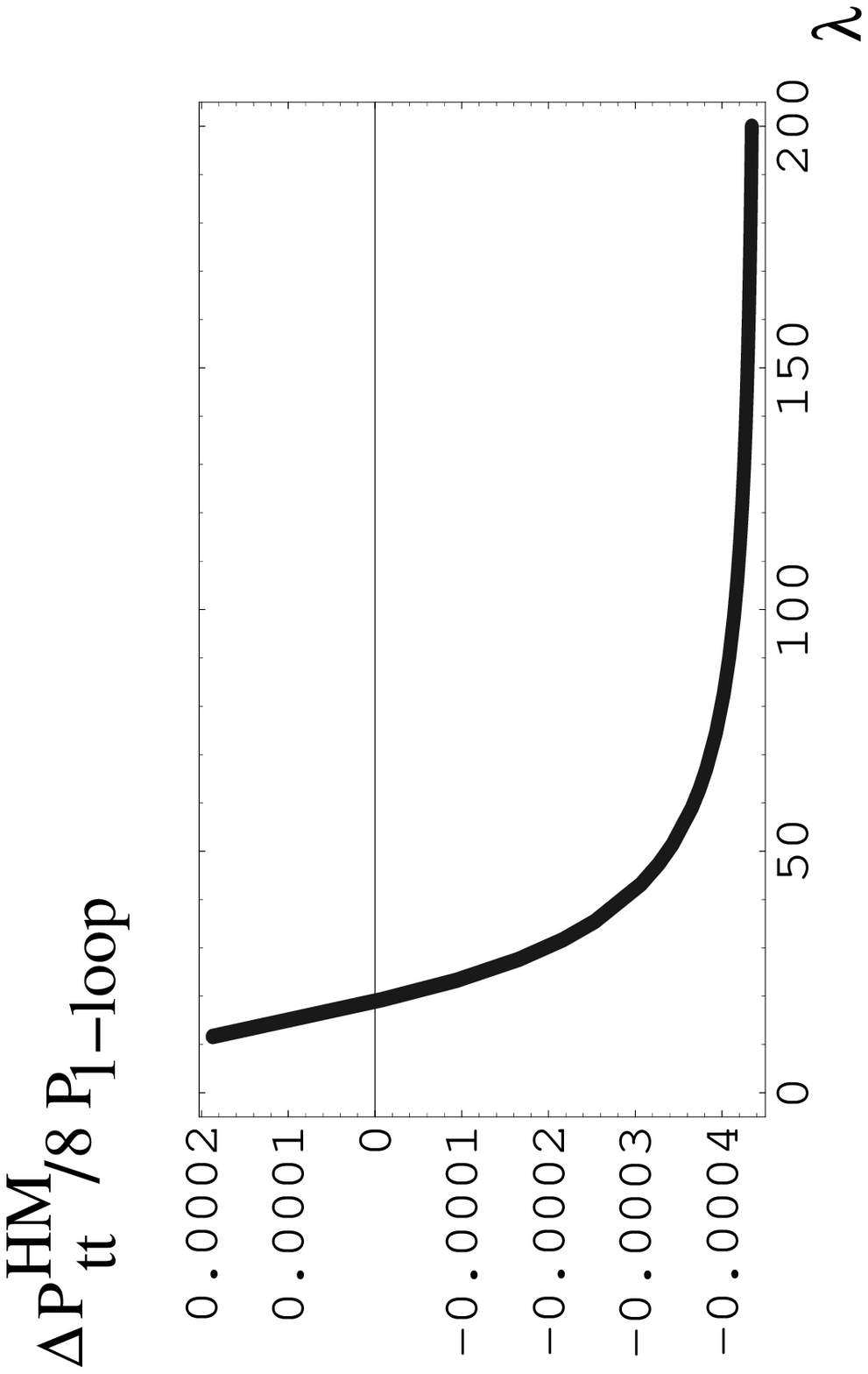}
\end{center}
\caption{Ratio of $\frac{1}{8}\Delta P^{HM}_{tt}$ and $P_{\tiny\mbox{1-loop}}$ as a function of $11.65\le\lambda\le 200$.\label{Fig-5.ps}}
\end{figure}

\begin{figure}
\begin{center}
\leavevmode
\leavevmode
\vspace{5.3cm}
\includegraphics{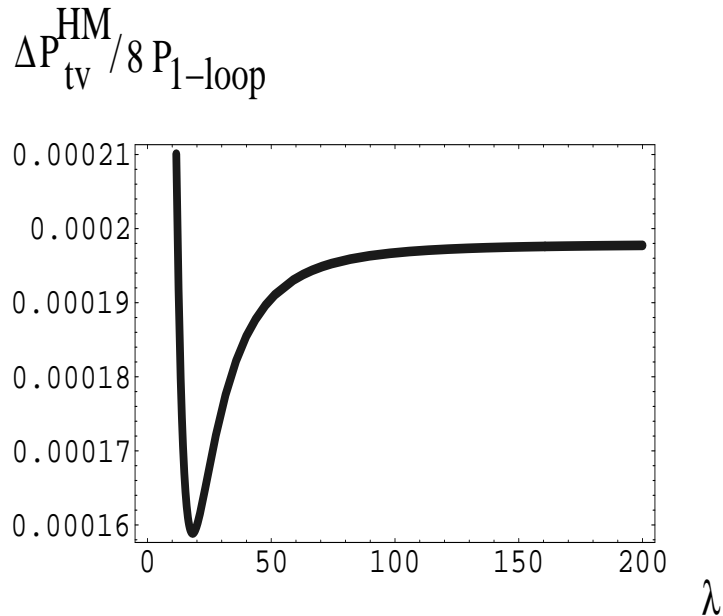}
\end{center}
\caption{An upper bound for the modulus of the ratio of $\frac{1}{8}\Delta P^{HM}_{tv}$ and 
$P_{\tiny\mbox{1-loop}}$ as a function of $11.65\le\lambda\le 200$.\label{Fig-6.ps}}
\end{figure}

\begin{figure}
\begin{center}
\leavevmode
\leavevmode
\vspace{5.3cm}
\includegraphics{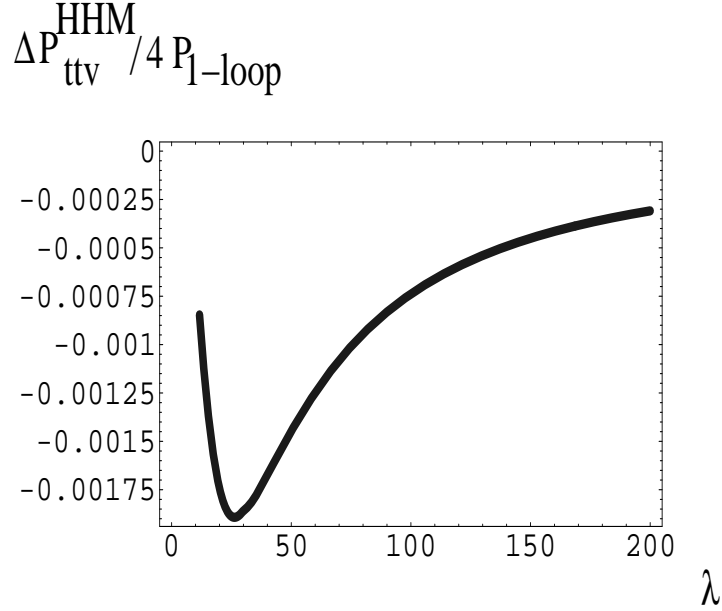}
\end{center}
\caption{Ratio of $\frac{1}{4}\Delta P^{HHM}_{ttv}$ and $P_{\tiny\mbox{1-loop}}$ 	as a function of $11.65\le\lambda\le 200$.\label{Fig-7.ps}}
\end{figure}

Having performed the numerical integrations, we now are in a position to 
present our results for each contributing diagram by plotting the 
ratio of two-loop to one-loop diagrams as a function of the dimensionless 
temperature $\lambda_c=11.65\le lambda\le 200$. 
Figs.\,\ref{Fig-4.ps} through \ref{Fig-7.ps} show the results. Notice that the one-loop result, 
see \cite{Hofmann2004} for a calculation, does not contain the 
contribution of the ground state. Notice also, that we kept $e\equiv 5.1$ 
for all values of $\lambda$ thus ignoring
the logarithmic blow-up of Eq.\,(\ref{logpole}). Due to the 
exponential suppression for large $e$ this yields an upper 
bound for the modulus of each diagram in the critical region.  
\newpage 
Our computation indicates that the two-loop corrections 
are at most 0.2\% of the one-loop result. The dominant contribution comes 
from the nonlocal diagram in Fig.\,\ref{fig:contribdiagr}. 

\section{Summary and Outlook\label{SO}}

Our results can be summarized as follows: The picture of almost noninteracting 
thermal quasiparticles that was underlying the one-loop evolution of the effective coupling constant $e$ in the 
electric phase of a thermalized SU(2) Yang-Mills theory is confirmed by the two-loop calculation of the 
thermodynamical pressure. The (tiny) modification of the one-loop evolution 
equation for $e$ due to two-loop effects will be 
investigated in \cite{HofmannRohrer2004}. On a mesoscopic level this modification 
can be understood in terms of scattering processes off magnetic 
monopoles whose core size becomes comparable to the typical wave length $T^{-1}$ of a 
TLM mode for $T\searrow T_{c}$ where $T_{c}=\frac{\lambda_c\Lambda}{2\pi}=\frac{11.65\,\Lambda}{2\pi}$ denotes the critical 
temperature for the $2^{\tiny\mbox{nd}}$ order transition to the magnetic phase. For $T\gg T_{c}$ the magnetic charge
of a monopole is too much smeared to be 'seen' by the TLM mode. This simple fact arises from the constancy of $e$ for
large temperatures and the core size or charge radius $R(T)$ of a monopole being approximately 
its inverse mass $M$ \cite{Hofmann2004} 
\eqb
R(T)\sim M^{-1}(T)\sim e\sqrt{\frac{2\pi T}{\Lambda^3}}\,.
\eqe
Thus the quick die-off of the two-loop correction to the pressure 
at large $T$ (compare with Figs.\,\ref{Fig-4.ps} through 
\ref{Fig-7.ps} and ignore the fact that our estimate for 
$\Delta P^{HM}_{tv}$, see Fig.\,\ref{Fig-6.ps}, is too rough for large $T$ due 
to the omission of the vertex constraint and that a small infrared effect survives for large $T$ in 
Fig.\,\ref{Fig-5.ps} due to the masslessness of the TLM mode).
The mechanical analogon for this situation is as follows: 
Imagine a box filled with heavy lead balls being at rest and light ping-pong balls moving around them.
Now, switch on an interaction between the two species (wavelength of TLM mode becomes comparable
to charge radius of monopole for $T\searrow T_c$). 
This will thermalize the system. However, the average momentum that is deprived from the ping-pong balls
and added to the lead balls does not have an effect on the partial thermodynamical pressure of the latter 
since their momenta only probe the exponential tail of their Bose distribution.
On the other hand, a decrease of the average ping-pong-ball momentum sizeably decreases their
partial thermodynamical pressure. This is seen in Fig.\,\ref{Fig-7.ps} by the (negative!) dip 
of the dominating two-loop correction.

Despite the large value of $e\sim 5.1$ the smallness of two-loop corrections emerges from the 
existence of compositeness constraints which in turn are derived 
from the existence of a nontrivial ground state. We expect no major complications 
when generalizing our computation to SU(N). The situation is somewhat reminiscent of ${\cal N}=2$ 
supersymmetric Yang-Mills theory where the {\sl perturbative} $\beta$ 
function for the gauge coupling is exact at one loop \cite{Seiberg1988}. 
The important conceptual difference is that the one-loop exactness in the supersymmetric 
case is inforced by a strong symmetry while in our approach to 
the ${\cal N}=0$ Yang-Mills theory the identification of the essential 
degrees of freedom makes the interactions thereof almost vanish. We expect 
that the loop expansion of the thermodynamical pressure of an SU(N) Yang-Mills theory 
is not asymptotic but converges very quickly.

An important application of our results arises: If the photon is generated by an SU(2) Yang-Mills 
theory of Yang-Mills scale $\Lambda\sim T_{\tiny\mbox{CMB}}\sim 10^{-4}\,$eV being at the boundary 
between the magnetic and electric phases but on the magnetic side\footnote{Only there is the photon precisely 
massless and completely unscreened: a situation which is dynamically stabilized by a dip of 
the energy density at $T_{\tiny\mbox{CMB}}$ \cite{Hofmann2004}.} then light, 
being released at the time of decoupling of the CMB (deep within the electric phase of SU(2)$_{\tiny\mbox{CMB}}$), 
must have travelled through a 'lattice' of scattering centers ({\sl dual} magnetic, that is, 
electrically charged monopoles) shortly before the Universe settled 
into the CMB dip where the monopoles are condensed into a classical field \cite{Hofmann2004}. This effect 
is seen in Fig.\,\ref{Fig-7.ps} by a decrease of the dominating two-loop correction 
to the pressure for $T$ approaching $T_c$ (that is $T_{\tiny\mbox{CMB}}$) from above. 
The observable effect should be a cosmic Laue diagram with a large quadrupole contribution 
and manifest itself in terms of a large-angle 'anomaly' in the power spectrum of temperature fluctuations in 
the cosmic microwave background. Such an 'anomaly' indeed has been reported 
by the WMAP collaboration \cite{WMAP2003}. 

\section*{Acknowledgments}

It is a pleasure to thank Alan Guth for a very stimulating 
discussion about the implications of SU(2)$_{\tiny\mbox{CMB}}$ 
for the CMB power spectrum. Useful conversations with Robert 
Brandenberger, John Moffat, Nucu Stamatescu, Dirk Rischke, and 
Frank Wilczek are gratefully acknowledged.  

\section*{Appendix}

Here we evaluate the contractions of the tensor structures as they 
appear in Eqs.\,(\ref{eqn:DP2local}) and (\ref{eqn:DP2nonlocal}). Exploiting Eq.\,(\ref{eqn:TLMfeatures}), 
the contractions for local contributions are:\\
(1) Local, TLH-TLH:
\begin{eqnarray}
&&\Gamma_{[4]abcd}^{\mu\nu\rho\sigma}\delta_{ab}\tilde{D}_{\mu\nu}(p)\delta_{cd}\tilde{D}_{\rho\sigma}(k)
=-ie^2[\epsilon_{abe}\epsilon_{cde}(g^{\mu\rho}g^{\nu\sigma}-g^{\mu\sigma}g^{\nu\rho})+\nonumber\\
&&\epsilon_{ace}\epsilon_{bde}(g^{\mu\nu}g^{\rho\sigma}-g^{\mu\sigma}g^{\nu\rho})
+\epsilon_{ade}\epsilon_{bce}(g^{\mu\nu}g^{\rho\sigma}-g^{\mu\rho}g^{\nu\sigma})]\times\nonumber\\
&&\delta_{ab}\left(g_{\mu\nu}-\frac{p_{\mu}p_{\nu}}{m^2}\right)\delta_{cd}\left(g_{\rho\sigma}-
\frac{k_{\rho}k_{\sigma}}{m^2}\right)\nonumber\\
&=&-ie^2\epsilon_{ace}\epsilon_{ace}\big[2g^{\mu\nu}g^{\rho\sigma}-g^{\mu\sigma}g^{\nu\rho}-g^{\mu\rho}g^{\nu\sigma}\big]\times\nonumber\\
&&\Big(g_{\mu\nu}-\frac{p_{\mu}p_{\nu}}{m^2}\Big)\Big(g_{\rho\sigma}-\frac{k_{\rho}k_{\sigma}}{m^2}\Big)\nonumber\\
&=&-2ie^2\left(24-6\frac{p^2}{m^2}-6\frac{k^2}{m^2}+2\frac{p^2k^2}{m^4}-2\frac{(pk)^2}{m^4}\right)\,.\label{eqn:HHcontraction}
\end{eqnarray}
(2) Local, TLH-TLM:
\begin{eqnarray}
&&\Gamma_{[4]abcd}^{\mu\nu\rho\sigma}\delta_{ab}\tilde{D}_{\mu\nu}(p)\delta_{cd}\bar{D}_{\rho\sigma}(k)
=-ie^2[\epsilon_{abe}\epsilon_{cde}(g^{\mu\rho}g^{\nu\sigma}-g^{\mu\sigma}g^{\nu\rho})+\nonumber\\
&&\epsilon_{ace}\epsilon_{bde}(g^{\mu\nu}g^{\rho\sigma}-g^{\mu\sigma}g^{\nu\rho})
+\epsilon_{ade}\epsilon_{bce}(g^{\mu\nu}g^{\rho\sigma}-g^{\mu\rho}g^{\nu\sigma})]\times\nonumber\\
&&\delta_{ab}\left(g_{\mu\nu}-\frac{p_{\mu}p_{\nu}}{m^2}\right)\delta_{cd}\bar{D}_{\rho\sigma}(k)\nonumber\\
&=&-ie^2\epsilon_{ace}\epsilon_{ace}\big[2g^{\mu\nu}g^{\rho\sigma}-g^{\mu\sigma}g^{\nu\rho}-g^{\mu\rho}g^{\nu\sigma}\big]\times\nonumber\\
&&\Big(g_{\mu\nu}-\frac{p_{\mu}p_{\nu}}{m^2}\Big)\bar{D}_{\rho\sigma}(k)\nonumber\\
&=&-2ie^2\left(-12+4\frac{p^2}{m^2}+2\frac{\vec{p}^2\sin^2{\theta}}{m^2}\right)\,.\label{eqn:HMcontraction}
\end{eqnarray}
In Eq.\,(\ref{eqn:HMcontraction}) $\theta$ denotes the angle between ${\bf p}$ and ${\bf k}$. 
For the  nonlocal diagram we obtain:
\begin{eqnarray}
&&\Gamma_{[3]abc}^{\lambda\mu\nu}(p,k,q)\Gamma_{[3]rst}^{\rho\sigma\tau}(p,k,q)\delta_{ar}\tilde{D}_{\lambda\rho}(p)
\delta_{bs}\tilde{D}_{\mu\sigma}(k)\delta_{ct}\bar{D}_{\nu\tau}(q)\nonumber\\
&=&e^2\epsilon_{abc}\epsilon_{rst}\big[g^{\lambda\mu}(p-k)^{\nu}+g^{\mu\nu}(k-q)^{\lambda}+g^{\nu\lambda}(q-p)^{\mu}\big]\times\nonumber\\
&&\big[g^{\rho\sigma}(p-k)^{\tau}+g^{\sigma\tau}(k-q)^{\rho}+g^{\tau\rho}(q-p)^{\sigma}\big]\times\nonumber\\
&&\delta_{ar}\delta_{bs}\delta_{ct}\Big( g_{\lambda\rho}-\frac{p_{\lambda}p_{\rho}}{m^2}\Big)
\Big( g_{\mu\sigma}-\frac{k_{\mu}k_{\sigma}}{m^2}\Big)
\bar{D}_{\nu\tau}(q)\,.\label{eqn:HHMuncontracted}
\end{eqnarray}
For not loosing track, we split the calculation into terms $\propto e^2$, $\propto\frac{e^2}{m^2}$ 
and $\propto\frac{e^2}{m^4}$ and keep $\bar{D}$ uncontracted in a first step. 
The contraction of structure constants $\epsilon_{ab3}\epsilon_{ab3}$ gives an additional factor 2.\\
Term $\propto 2e^2$:
\begin{eqnarray}
&&\big[g^{\lambda\mu}(p-k)^{\nu}+g^{\mu\nu}(k-q)^{\lambda}
+g^{\nu\lambda}(q-p)^{\mu}\big]
\big[g^{\rho\sigma}(p-k)^{\tau}+\nonumber\\
&&g^{\sigma\tau}(k-q)^{\rho}+g^{\tau\rho}(q-p)^{\sigma}\big] g_{\lambda\rho}g_{\mu\sigma}\bar{D}_{\nu\tau}(q)
\nonumber\\
&=&\big[g_{\rho\sigma}(p-k)^{\nu}+g_{\sigma}^{\nu}(k-q)_{\rho}+g^{\nu}_{\rho}(q-p)_{\sigma}\big]\times\nonumber\\
&&\big[g^{\rho\sigma}(p-k)^{\tau}+g^{\sigma\tau}(k-q)^{\rho}+g^{\tau\rho}(q-p)^{\sigma}\big]\bar{D}_{\nu\tau}(q)
\nonumber\\
&=&\big[4(p-k)^{\nu}(p-k)^{\tau}+(p-k)^{\nu}(k-q)^{\tau}+(p-k)^{\nu}(q-p)^{\tau}+\nonumber\\
&&(k-q)^{\nu}(p-k)^{\tau}+(k-q)^2g^{\nu\tau}+(q-p)^{\nu}(k-q)^{\tau}+\nonumber\\
&&(q-p)^{\nu}(p-k)^{\tau}+(k-q)^{\nu}(q-p)^{\tau}+(q-p)^2g^{\nu\tau}\big]\bar{D}_{\nu\tau}(q)\nonumber\\
&=&\big[2p^{\nu}p^{\tau}+2k^{\nu}k^{\tau}-6p^{\nu}k^{\tau}+(q-p)^2g^{\nu\tau}+(k-q)^2g^{\nu\tau}\big]
\bar{D}_{\nu\tau}(q)\nonumber\\
&=&2\Big(\vec{p}^2-\frac{(\vec{p}\vec{q})^2}{|\vec{q}|^2}\Big)+
2\Big(\vec{k}^2-\frac{(\vec{k}\vec{q})^2}{|\vec{q}|^2}\Big)
-6\Big(\vec{p}\vec{k}-\frac{(\vec{p}\vec{q})(\vec{k}\vec{q})}{|\vec{q}|^2}\Big)-\nonumber\\
&&2(q-p)^2-2(k-q)^2\,.\label{eqn:HHMcontraction1}
\end{eqnarray}
Terms proportional to $q^{\nu}$ or $q^{\tau}$ have been omitted after the second-last equal sign in 
Eq.\,(\ref{eqn:HHMcontraction1}) because, when 
contracted with $\bar{D}_{\nu\tau}(q)$, they vanish. Again, 
using Eq.\,(\ref{eqn:TLMfeatures}) the expression after the last equal sign in 
Eq.\,(\ref{eqn:HHMcontraction1}) easily follows.\\ 
Next we look at the two terms proportional to $2\frac{e^2}{m^2}$ 
(compare with Eq.\,(\ref{eqn:HHMuncontracted})).\\
The first one is: 
\begin{eqnarray}
\label{eqn:HHMcontraction2}
&&\big[g^{\lambda\mu}(p-k)^{\nu}+g^{\mu\nu}(k-q)^{\lambda}
+g^{\nu\lambda}(q-p)^{\mu}\big]
\big[g^{\rho\sigma}(p-k)^{\tau}+\nonumber\\
&&g^{\sigma\tau}(k-q)^{\rho}+g^{\tau\rho}(q-p)^{\sigma}\big]g_{\lambda\rho}k_{\mu}k_{\sigma}\bar{D}_{\nu\tau}(q)
\nonumber\\
&=&\big[k_{\rho}k_{\sigma}(p-k)^{\nu}+k^{\nu}k_{\sigma}(k-q)_{\rho}+k(q-p)g^{\nu}_{\rho}k_{\sigma}\big]\times\nonumber\\
&&\big[g^{\rho\sigma}(p-k)^{\tau}
+g^{\sigma\tau}(k-q)^{\rho}+g^{\tau\rho}(q-p)^{\sigma}\big]\bar{D}_{\nu\tau}(q)\nonumber\\
&=&\big[k^2(p-k)^{\nu}(p-k)^{\tau}+k(k-q)(p-k)^{\nu}k^{\tau}+k(q-p)(p-k)^{\nu}k^{\tau}+\nonumber\\
&&k(k-q)k^{\nu}(p-k)^{\tau}+(k-q)^2k^{\nu}k^{\tau}+k(q-p)k^{\nu}(k-q)^{\tau}+\nonumber\\
&&k(q-p)k^{\nu}(p-k)^{\tau}+k(q-p)(k-q)^{\nu}k^{\tau}+[k(q-p)]^2g^{\nu\tau}\big]\bar{D}_{\nu\tau}(q)\nonumber\\
&=&[k^2p^{\nu}p^{\tau}+q^2k^{\nu}k^{\tau}-2(kp) p^{\nu}k^{\tau}+[k(q-p)]^2g^{\nu\tau}\big]\bar{D}_{\nu\tau}(q)
\nonumber\\
&=&k^2\Big(\vec{p}^2-\frac{(\vec{p}\vec{q})^2}{|\vec{q}|^2}\Big)+q^2\Big(\vec{k}^2-\frac{(\vec{k}\vec{q})^2}{|\vec{q}|^2}\Big)
-2pk\Big(\vec{p}\vec{k}-\frac{(\vec{p}\vec{q})(\vec{k}\vec{q})}{|\vec{q}|^2}\Big)\nonumber\\
&&-2[k(q-p)]^2\,.
\end{eqnarray}
The second term $\propto 2\frac{e^2}{m^2}$ either is obtained by a 
direct calculation or by just
exchanging $p\leftrightarrow k$ in Eq.\,(\ref{eqn:HHMcontraction2}):
\begin{eqnarray}
&&\big[g^{\lambda\mu}(p-k)^{\nu}+g^{\mu\nu}(k-q)^{\lambda}
+g^{\nu\lambda}(q-p)^{\mu}\big]\big[g^{\rho\sigma}(p-k)^{\tau}+\nonumber\\
&&g^{\sigma\tau}(k-q)^{\rho}+g^{\tau\rho}(q-p)^{\sigma}\big] g_{\mu\sigma}p_{\lambda}p_{\rho}\bar{D}_{\nu\tau}(q)
\nonumber\\
&=&p^2\Big(\vec{k}^2-\frac{(\vec{k}\vec{q})^2}{|\vec{q}|^2}\Big)
+q^2\Big(\vec{p}^2-\frac{(\vec{p}\vec{q})^2}{|\vec{q}|^2}\Big)
-2pk\Big(\vec{p}\vec{k}-\frac{(\vec{p}\vec{q})(\vec{k}\vec{q})}{|\vec{q}|^2}\Big)-\nonumber\\
&&2[p(q-k)]^2\,.\label{eqn:HHMcontraction3}
\end{eqnarray}
Finally, the term $\propto 2\frac{e^2}{m^4}$ is given by
\begin{eqnarray}
&&\big[g^{\lambda\mu}(p-k)^{\nu}+g^{\mu\nu}(k-q)^{\lambda}
+g^{\nu\lambda}(q-p)^{\mu}\big]
\big[g^{\rho\sigma}(p-k)^{\tau}+\nonumber\\
&&g^{\sigma\tau}(k-q)^{\rho}+g^{\tau\rho}(q-p)^{\sigma}\big] p_{\lambda}p_{\rho}k_{\mu}k_{\sigma}
\bar{D}_{\nu\tau}(q)\nonumber\\
&=&\big[(pk)p_{\rho}k_{\sigma}(p-k)^{\nu}+p(k-q)p_{\rho}k_{\sigma}k^{\nu}+k(q-p)p_{\rho}k_{\sigma}p^{\nu}\big]\times\nonumber\\
&&\big[g^{\rho\sigma}(p-k)^{\tau}+g^{\sigma\tau}(k-q)^{\rho}+g^{\tau\rho}(q-p)^{\sigma}\big]\bar{D}_{\nu\tau}(q)
\nonumber\\
&=&\big[(pk)(p-k)^{\nu}+[p(k-q)]k^{\nu}+[k(q-p)]p^{\nu}\big]\nonumber\\
&&\big[(pk)(p-k)^{\tau}+[p(k-q)]k^{\tau}+[k(q-p)]p^{\tau}\big]\bar{D}_{\nu\tau}(q)\nonumber\\
&=&\big[(kq)^2p^{\nu}p^{\tau}+(pq)^2k^{\nu}k^{\tau}-2(pq)(kq)p^{\nu}k^{\tau}\big]\bar{D}_{\nu\tau}(q)\nonumber\\
&=&(kq)^2\Big(\vec{p}^2-\frac{(\vec{p}\vec{q})^2}{|\vec{q}|^2}\Big)
+(pq)^2\Big(\vec{k}^2-\frac{(\vec{k}\vec{q})^2}{|\vec{q}|^2}\Big)-\nonumber\\
&&2(pq)(kq)\Big(\vec{p}\vec{k}-\frac{(\vec{p}\vec{q})(\vec{k}\vec{q})}{|\vec{q}|^2}\Big)\,.\label{eqn:HHMcontraction4}
\end{eqnarray}
Now, adding up Eqs.(\ref{eqn:HHMcontraction1}) through (\ref{eqn:HHMcontraction4}) (taking care of the 
correct signs), we have
\begin{eqnarray}
\mbox{Eq.(\ref{eqn:HHMuncontracted})}&=&2e^2\left[\mbox{Eq.(\ref{eqn:HHMcontraction1})}-\mbox{Eq.(\ref{eqn:HHMcontraction2})}
-\mbox{Eq.(\ref{eqn:HHMcontraction3})}+\mbox{Eq.(\ref{eqn:HHMcontraction4})}\right]\nonumber\\
&=&2e^2\Big\{2\frac{[p(q-k)]^2}{m^2}+\frac{2[k(q-p)]^2}{m^2}-2(q-p)^2-2(k-q)^2+\nonumber\\
&&[2-\frac{k^2}{m^2}-\frac{q^2}{m^2}+\frac{(kq)^2}{m^4}]\Big(\vec{p}^2-\frac{(\vec{p}\vec{q})^2}{|\vec{q}|^2}\Big)+\nonumber\\
&&[2-\frac{q^2}{m^2}-\frac{p^2}{m^2}+\frac{(pq)^2}{m^4}]\Big(\vec{k}^2-\frac{(\vec{k}\vec{q})^2}{|\vec{q}|^2}\Big)-\nonumber\\
&&[6-4\frac{pk}{m^2}+2\frac{(pq)(kq)}{m^4}]\Big(\vec{p}\vec{k}-\frac{(\vec{p}
\vec{q})(\vec{k}\vec{q})}{|\vec{q}|^2}\Big)\Big\}\,.
\end{eqnarray}
Using momentum conservation at the vertices, that is $q=-p-k$, we find:
\begin{eqnarray}
\Big(\vec{p}^2-\frac{(\vec{p}\vec{q})^2}{|\vec{q}|^2}\Big)=\Big(\vec{k}^2-\frac{(\vec{k}\vec{q})^2}{|\vec{q}|^2}\Big)
=-\Big(\vec{p}\vec{k}-\frac{(\vec{p}\vec{q})(\vec{k}\vec{q})}{|\vec{q}|^2}\Big)=
\frac{\vec{p}^2\vec{k}^2 \sin^2{\theta}}{(\vec{p}+\vec{k})^2}\,.
\end{eqnarray}
And thus,\\
(3) Nonlocal, TLH-TLH-TLM:
	\begin{eqnarray}
	&&\Gamma_{[3]abc}^{\lambda\mu\nu}(p,k,q)\Gamma_{[3]abc}^{\rho\sigma\tau}(-p,-k,-q)
	\tilde{D}_{\lambda\rho}(p)\tilde{D}_{\mu\sigma}(k)\bar{D}_{\nu\tau}(q)\nonumber\\
	&=&2e^2\Big[10p^2+10k^2+16pk-2\frac{k^4}{m^2}-2\frac{p^4}{m^2}-8\frac{p^2(pk)}{m^2}-8\frac{k^2(pk)}{m^2}-\nonumber\\
	&&16\frac{(pk)^2}{m^2}-\frac{\vec{p}^2\vec{k}^2\sin^2{\theta}}{(p+k)^2}\Big(10-3\frac{p^2}{m^2}-3\frac{k^2}{m^2}
	-8\frac{pk}{m^2}+\frac{p^4}{m^4}+\nonumber\\
	&&\frac{k^4}{m^4}+4\frac{p^2(pk)}{m^4}+4\frac{k^2(pk)}{m^4}
	+4\frac{(pk)^2}{m^4}+2\frac{p^2k^2}{m^4}\Big)\Big]\label{eqn:HHMcontracted}\,.
	\end{eqnarray}
Here, we have used the fact that $\Gamma(-p,-k,-q)=-\Gamma(p,k,q)$.

\baselineskip25pt
\end{document}